# CMOS-Compatible Ising Machines built using Bistable Latches Coupled through Ferroelectric Transistor Arrays

Antik Mallick[1], Zijian Zhao[2], Mohammad Khairul Bashar[1], Shamiul Alam[3], Md Mazharul Islam[3], Yi Xiao[4], Yixin Xu[4], Ahmedullah Aziz[3], Vijaykrishnan Narayanan[4], Kai Ni[2], Nikhil Shukla[1]*

[1]Department of Electrical and Computer Engineering, University of Virginia, Charlottesville, VA 22904, USA

[2]Department of Electrical and Microelectronic Engineering, Rochester Institute of Technology, Rochester, NY 14623, USA.

[3] Department of Electrical Engineering and Computer Science, University of Tennessee, Knoxville, TN 37996, USA

[4]Department of Computer Science & Engineering, Pennsylvania State University, State College, PA 16801, USA

*email: ns6pf@virginia.edu

**Realizing compact and scalable Ising machines that are compatible with CMOS-process technology is crucial to the effectiveness and practicality of using such hardware platforms for accelerating computationally intractable problems. Besides the need for realizing compact Ising spins, the implementation of the coupling network, which describes the spin interaction, is also a potential bottleneck in the scalability of such platforms. Therefore, in this work, we propose an Ising machine platform that exploits the novel behavior of compact bi-stable CMOS-latches (cross-coupled inverters) as classical Ising spins interacting through highly scalable and CMOS-process compatible ferroelectric-$HfO_2$-based Ferroelectric FETs (FeFETs) which act as coupling elements. We experimentally demonstrate the prototype building blocks of this system, and evaluate the behavior of the scaled system using simulations. We project that the proposed architecture can compute Ising solutions with an efficiency of ~$1.04 \times 10^8$ solutions/W/second. Our work not only provides a pathway to realizing CMOS-compatible designs but also to overcoming their scaling challenges.**

Ising Machines, as dynamical systems, have recently shown promise for accelerating computationally challenging problems in combinatorial optimization. The intrinsic energy minimization in the highly interconnected system gives rise to rich spatio-temporal properties, which can subsequently be mapped to the solutions of many computationally intractable optimization problems[1,2]. However, the highly interconnected nature of the system also poses a significant implementation and scalability challenge for Ising platforms. In fact, the number of coupling elements (representing edges) required for mapping an arbitrary graph scales up quadratically (~$N^2$) with the number of nodes in the graph. Consequently, scaling the system to large sizes continues to be a significant challenge for most Ising machine designs. Our approach to addressing this challenge relies on developing novel hardware components that are not only compact but can also leverage the maturity of CMOS-process technology and integration.

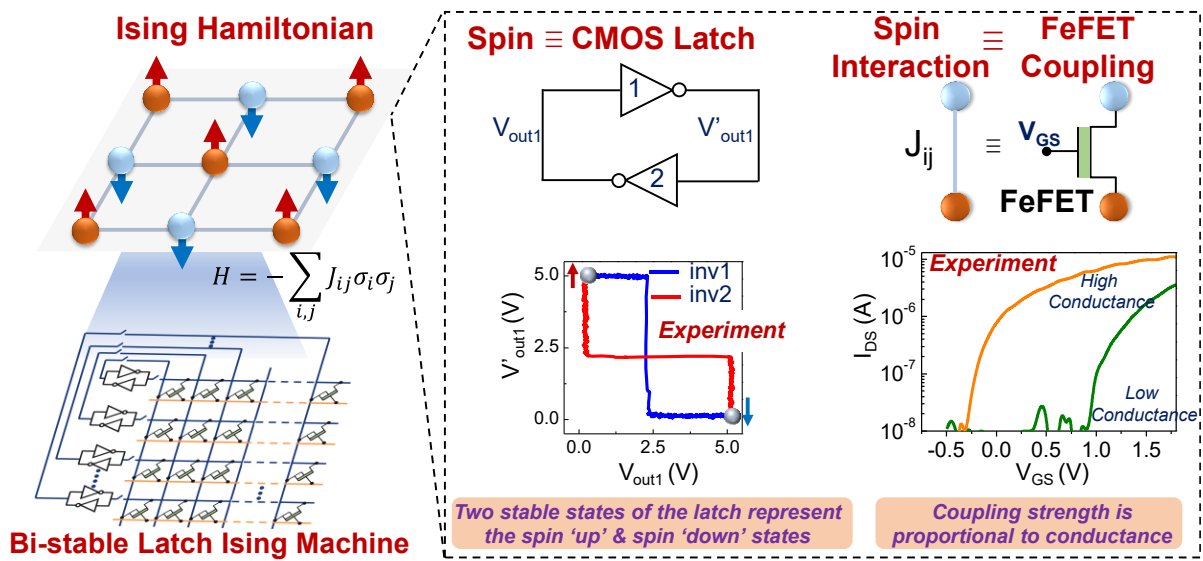

**Fig. 1. Overview of Bi-stable Latch based Ising Machine Hardware.** Proposed design for the Ising machine using CMOS latches (cross-coupled inverters) as artificial Ising spins. The interaction among the spins is implemented using a CMOS-process compatible FeFET based array.

There is a full pallet of hardware technologies[3-16], each with their advantages and shortcomings, that have been considered for implementing Ising machines. Quantum annealing approaches[17-21] using qubits offer the possibility of exponential speed up (by overcoming the fundamental hardness of the problem) but require cryogenic cooling. Besides cost, this requirement also restricts the type of applications where this approach would be practical. At the classical end, Ising machine implementations can be classified into the optical domain - using optoelectronic oscillators to design Coherent Ising Machines (CiM)[22-25], and the electronic domain using a variety of classical spin implementations. CiMs offer advantages such as speed as well as a relatively broad dynamic range for the implementation of weights[26]. However, such implementations have traditionally been bulky, requiring long optical fibers, although there is some recent work on monolithic integration[27-29].

Electronic Ising machines have relied on the following approaches: iterative annealing in memory (AIM)[30,31], that digitally emulates the Ising model[32,33] and the actual implementation of the dynamical system[34-38]. While the former approach essentially minimizes the Ising energy using a heuristic iterative approach, the latter method (relevant to this work) uses the hardware as classical spins and maps the energy minimization in the hardware directly to the minimization of the Ising Hamiltonian. Various devices such as oscillators[39-46] and ZIV diodes[47] have been experimentally demonstrated as classical Ising spins. More recently, CMOS-based (bi-stable) latches have also been *theoretically* shown to behave as Ising spins as well[48]. Here, we experimentally demonstrate CMOS-latches as highly scalable and compact Ising spins. Additionally, in all of the electronic designs, the implementation of the coupling network continues to be a significant challenge for scaling. To address this, we propose to exploit the non-volatile behavior of CMOS-compatible FeFET memory arrays (in fact, the FeFETs used in this work are built in 28 nm high-κ metal gate technology platform) to implement the interaction among the spins (CMOS latches). Consequently, our work enables a pathway to a compact Ising platform (Fig. 1) that is positioned to exploit the maturity of CMOS process technology to realize a scalable solution.

**Results**

**CMOS latch as a classical spin.** We first focus on experimentally evaluating the behavior of CMOS latches as classical spins with simple resistive coupling. The theoretical foundation for the latch-based Ising machine was elegantly formulated by J. Roychowdhury[48] wherein the energy function for a resistively coupled system of latches was shown to map to the Ising Hamiltonian. Fig. 2 shows the experimentally observed behavior of a system of two latches with positive ($J_{ij} = +1$) and negative ($J_{ij} = -1$) coupling over 100 runs; the details of the setup used in the experiments are discussed in supplementary note 1. It can be observed that the latches settle to the same voltage output level (i.e., the read terminals have the same outputs (0,0) or ($V_{DD}$, $V_{DD}$), when positively coupled, and opposite voltage levels (i.e., the read terminals have opposite outputs ($V_{DD}$,0) or (0, $V_{DD}$)), when negatively coupled. We note, however, that the exact output (i.e., whether $V_{out1}$ or $V_{out2}$ settles to $V_{DD}$ or 0) shows statistical behavior, as expected.

Building on this coupled two-latch spin system, we evaluate the dynamics of a system of four negatively coupled latches as shown in Fig. 3a-c. Our choice of negative coupling is motivated by the fact that the dynamics of such an Ising network can be directly mapped to computing the Maximum Cut (MaxCut) of a topologically equivalent graph- the MaxCut of a graph is defined as the challenge of dividing the nodes of a graph into two sets such that the number of common edges is maximized (unweighted graphs are considered here).

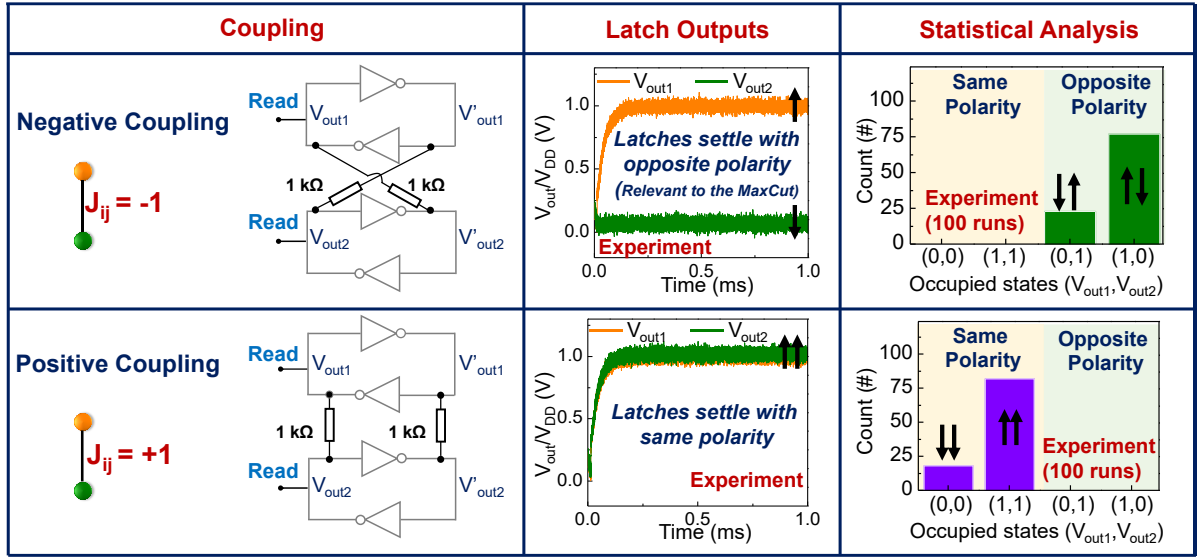

**Fig. 2. Coupling of two CMOS latch-based artificial Ising spins.** Negative ($J_{ij}$=-1) and positive coupling ($J_{ij}$=+1) among the latches. The latch outputs (at the read terminal) always settle to the opposite (same) polarity when the negatively (positively) coupled, respectively, when the system is powered up. It is noted though that the exact output (i.e., whether $V_{out1}$ or $V_{out2}$ settle to 1 (=$V_{DD}$) or 0 (=0V)) shows probabilistic behavior.

When powered up, the latch outputs ($L_1$=$L_3$=$V_{DD}$ (+1); $L_2$=$L_4$=0V (-1)) correspond to the two sets created by the MaxCut (Fig.3b) and yield an optimal MaxCut solution of 4 (Fig. 3c). Furthermore, to rule out the possibility that the output states of the latches are resulting from any inherent asymmetry among them, we map the nodes of the graph in Fig. 3 to different physical latches. The results, discussed in supplementary note 2, show that optimal solutions are observed in all the cases irrespective of the mapping, implying that the observed relationships between the latch outputs arise from the interaction among the latches governed by the minimization of the energy of the system. We also evaluate the effect of the coupling strength on the system properties by tuning the coupling resistance ($J_{ij} \propto 1/R$). It can be observed that the system exhibits the desired functionality as an Ising machine only in a limited coupling range (Fig. 3d). When the coupling is very weak (large R), the latch outputs are essentially decoupled (to a varying degree depending on the coupling strength) resulting in sub-optimal 'solutions'. In contrast, when the coupling is very strong (small R), the latches settle to a nearly common state (~$V_{DD}$/2). Next, we experimentally evaluate the MaxCut on graphs of up to 10 nodes. Fig. 3g shows the stochastic behavior of the system (expected for the Ising machine) for a representative graph with 10 nodes (Fig. 3e) measured over 150 iterations. Optimal solutions are observed in 86 out of the 150 measurements (Fig. 3h). The sub-optimal solutions in the remaining measurements result from the system getting trapped in local minima in the high dimensional phase space (Fig. 3f). Further, we also experimentally compute the MaxCut on multiple graph configurations (=20) up to 10 nodes (Fig. 3i); each graph is measured 10 times. Optimal MaxCut solutions are obtained in 17 of the 20 graphs.

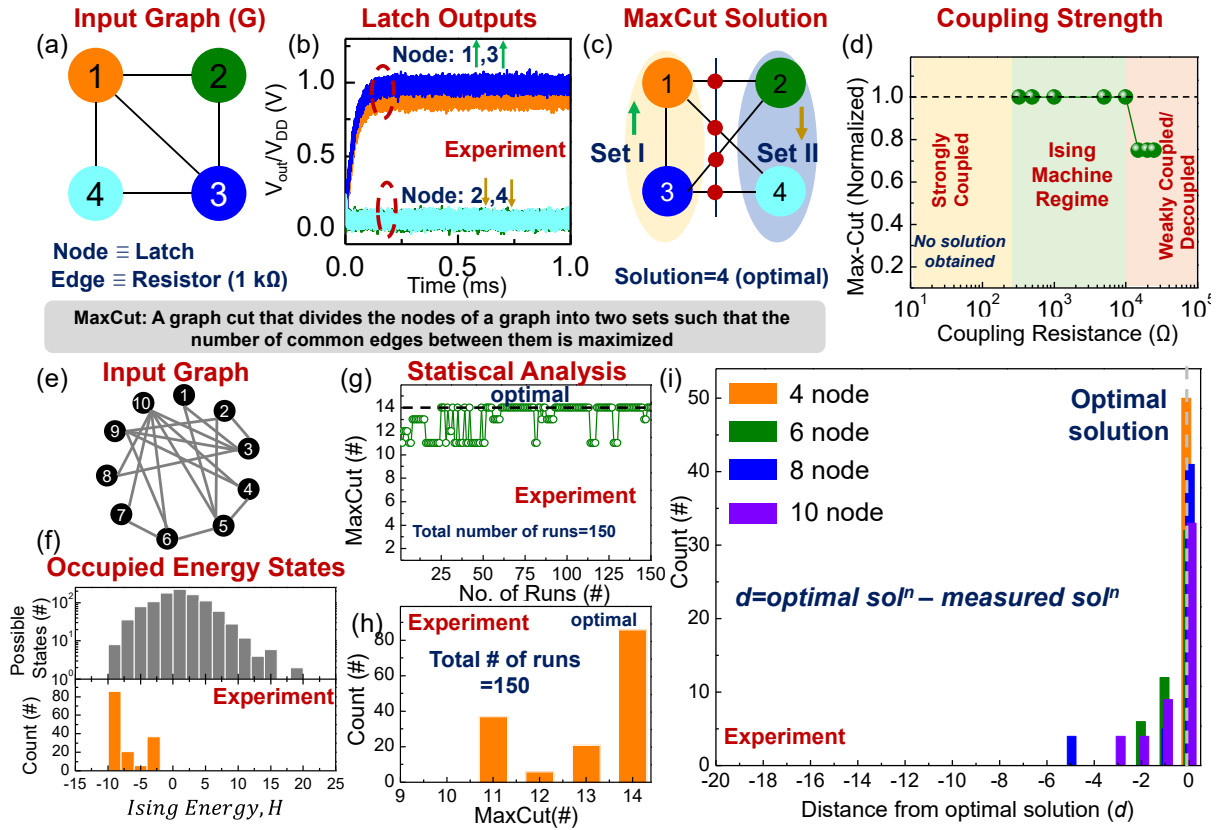

**Fig. 3. Solving MaxCut using Latch based Ising Machine.** (a) A representative 4 node graph problem. (b) Time domain outputs of the negative coupled ($J_{ij}$=-1) CMOS latches for the input graph. (c) MaxCut solution measured using the latch outputs in (b). (d) Effect of coupling resistance on the solution quality. (e) A representative network of 10 spins with randomly generated interactions (represented by edges). (f) Distribution showing occupied energy states (represented by H) and their frequency (orange) compared to the complete solution space (grey) for the problem. Experimentally observed MaxCut solution over 150 trials (g), illustrating the statistical behavior of the Ising machine (h). (i) Experimentally measured MaxCut solutions measured for multiple graphs of various size from 4-10 nodes. Each graph is measured 10 times, and the measured solution is represented by its distance *d* from the optimal solution.

**Coupling spins using FeFETs.** While the above experiments showcase the functional behavior of CMOS latches as classical Ising spins, the implementation of programmable monolithic resistors as coupling elements can be challenging and area inefficient, particularly in scaled systems. We therefore evaluate, first at a singular device level, the possibility of using non-volatile FeFETs (Fig. 4a) as programmable coupling elements between the CMOS latches. Our choice of using the FeFET as the coupling element is motivated by the fact that FeFETs are compatible with CMOS process technology, provide a wide dynamic range for the resistance (coupling strength), and can be efficiently integrated and programmed in a scalable array that is required to map the spin interactions in the entire network. We envision that the tunable threshold voltage of the FeFET would allow us to program the interaction between the latches; the low $V_T$ (high conductance) state would correspond to $J_{ij}=\pm 1$ whereas the high $V_T$ (low conductance) state would correspond to $J_{ij}=0$. Fig. 4b shows the experimentally measured transfer characteristics of the ferroelectric-$HfO_2$-based FeFETs used in this work; the devices are fabricated (see methods) on 28 nm high-κ metal gate technology platform, as shown in the cross-sectional TEM image[49,50]. It features a doped HfO$_2$ layer as the ferroelectric and SiO$_2$ as the interlayer. Detailed processing information is described elsewhere[50]. Fig. 4c shows the memory window vs. programming voltage characteristics for the FeFET. When a programming voltage of ±4 V is used to program the FeFET state, a 100× modulation in the current is obtained for V$_{GS}$= 1V.

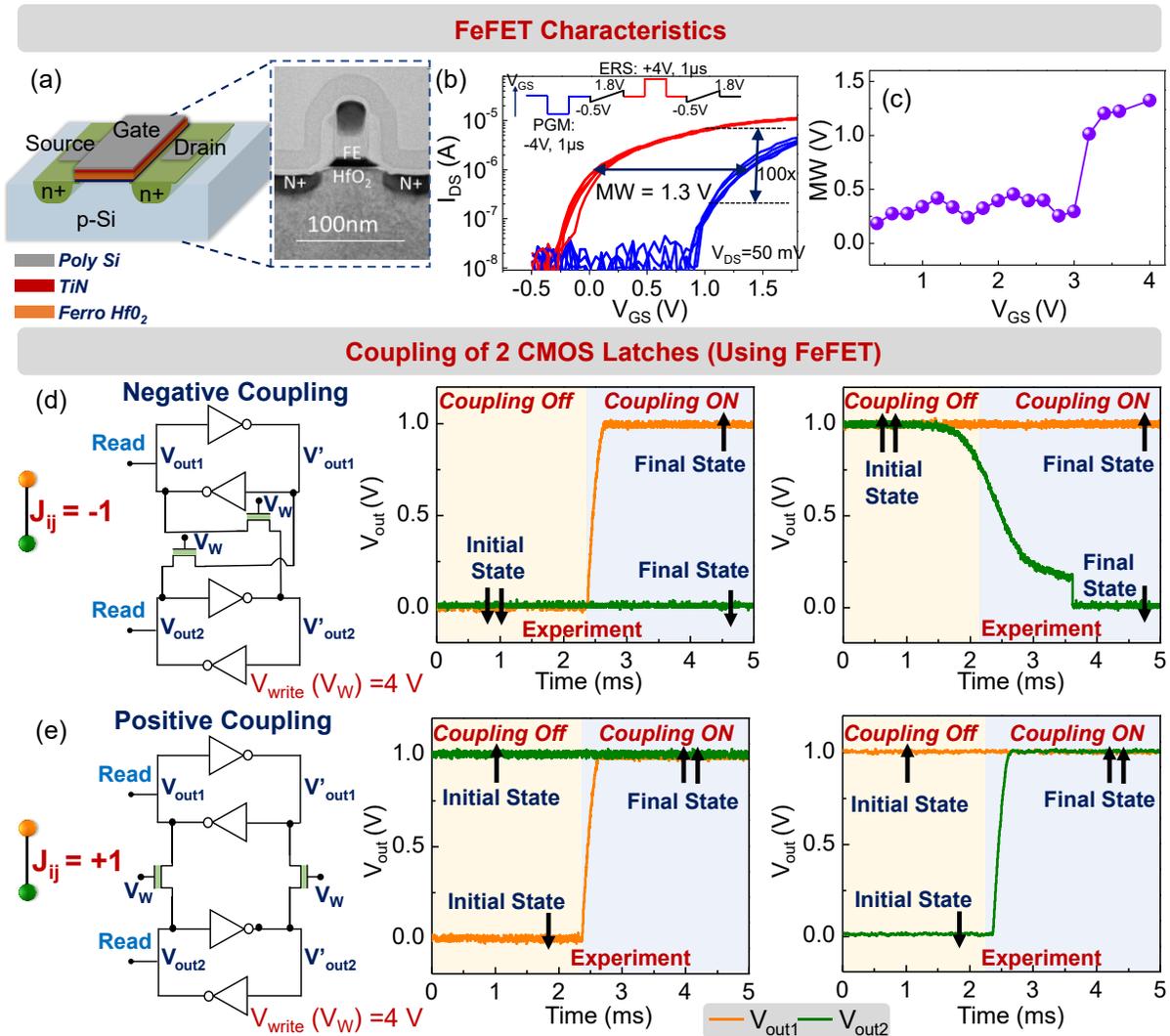

**Fig. 4. FeFET coupled CMOS latches.** (a) Schematic and TEM cross-section[50] of a 28 nm high-κ metal gate FeFET device. (b) I$_{DS}$–V$_{GS}$ characteristics of the FeFET (W/L=0.5/0.5 μm) after program and erase pulses. (c) Evolution of memory window (MW) as a function of write voltage (V$_{GS}$). FeFET coupled two-latch system settles (d) out-of-phase and (e) in-phase when the coupling is negative (J$_{ij}$=-1) and positive (J$_{ij}$=+1), respectively.

We subsequently characterize the behavior of the FeFET as a programmable coupling element in a two-latch system. To evaluate this, the FeFETs are first programmed into the low $V_T$/high conductance state ($J_{ij} = \pm 1$) using a programming pulse of magnitude +4V and a period of 1 µs. We test the interaction induced by the FeFETs among the latches by intentionally programming them into the 'incorrect' state, in order to observe the system evolve into the correct state i.e., when $J_{ij}$=-1 (+1), the latches are initialized into the same (opposite) states (0/$V_{DD}$), and subsequently, it is observed whether the system evolves to the correct state. During the initialization of the latches, the FeFETs are maintained at $V_{GS}$=-0.5V. This reduces the conductance of the FeFETs without affecting the threshold voltage. After the latches are initialized, the gate voltage is increased to $V_{GS}$ = 1.5 V, and the corresponding dynamics of the FeFET coupled CMOS latches are evaluated. We note that $V_{GS}$=1.5V was required to achieve the desired level of conductance from the FeFET-based coupling element (coupling strength) since the threshold voltage of the device has not optimized for this application. The gate voltage can be reduced to zero by appropriately adjusting the threshold as considered in the following simulations. Fig. 4d,e shows the observed behavior of the coupled system. Similar to the resistive coupling above, the coupled two-latch system settles in-phase (out-of-phase) when the coupling is positive $J_{ij}$=+1 (negative, $J_{ij}$=-1), respectively.

Using the above building blocks, we now explore a pathway to design a scalable Ising machine using latches as classing Ising spins and the FeFETs are programmable coupling elements. We propose to use a FeFET-based array to realize the coupling network among the latches (Fig. 5a). In this architecture, the rows (bit lines; BL) and the columns (source lines; SL) are connected to the drain and the source of the FeFET, respectively. Each row and column is driven by the two complementary outputs of a latch, effectively realizing the negative coupling ($J_{ij}$=-1), required to solve the MaxCut problem using the Ising model (Fig. 5a). We also note that the latches can be decoupled from the rows and columns using the transmission gate-based switches. This is required for the two-stage operation of the array as illustrated further. The word-line (WL), connected to the gates of all the FeFETs in a row, is used to program the FeFET state according to the adjacency matrix of the input graph.

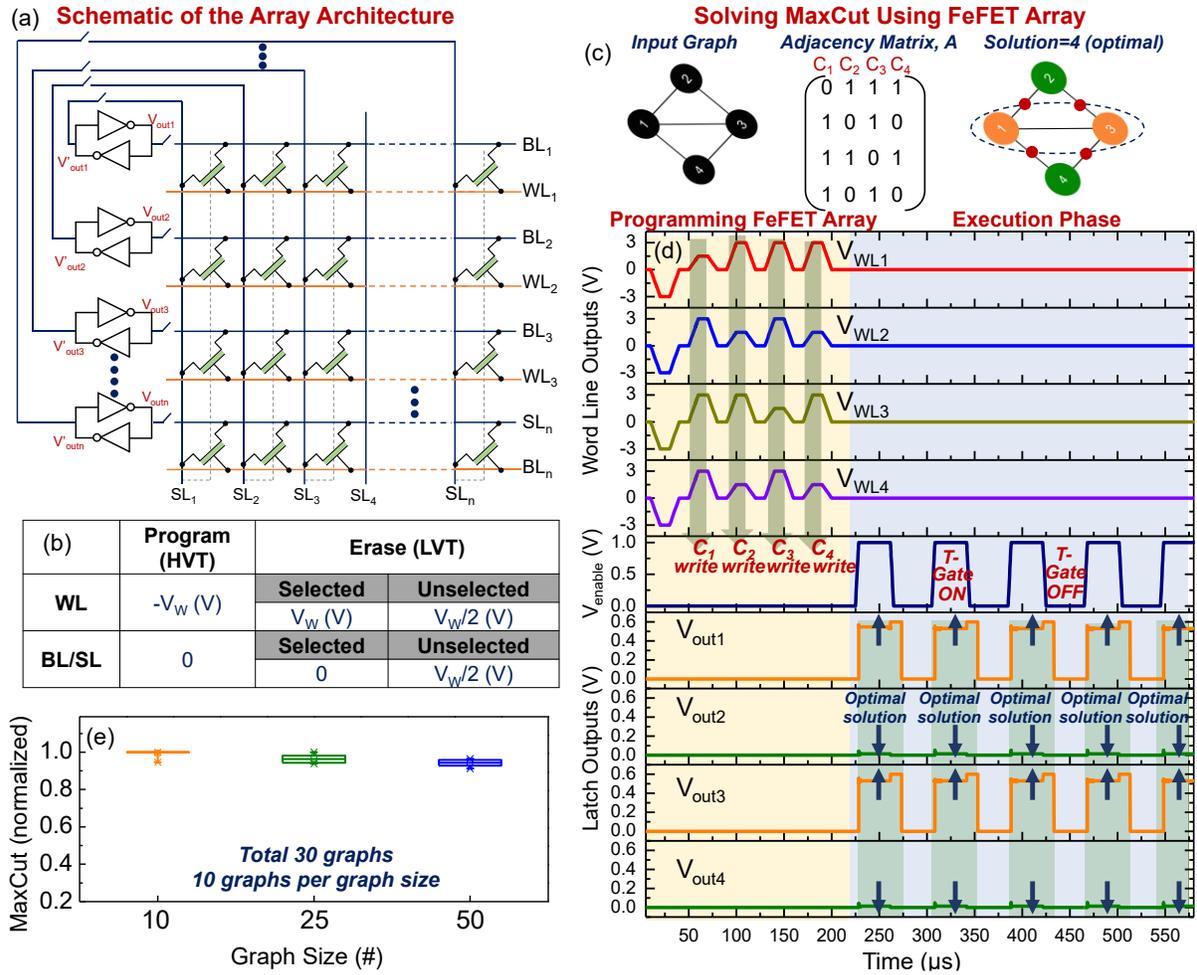

**Fig. 5. CMOS Latch-based Ising Machine with FeFET-based Coupling.** (a) Schematic of the FeFET array used to implement the coupling network and its interfacing with the CMOS-latches; negative coupling ($J_{ij}$=-1) is implemented here. (b) Half select $V_W/2$ write scheme that is used to program the FeFET array. (c) A representative 4 node graph problem. (d) Time domain output of the write voltages and latch outputs for solving the representative problem in (c). (e) MaxCut solutions obtained for graphs of various size up to 50 nodes. The graphs were randomly generated; 30 graphs were tested in total (10 different graph configurations per node).

The above array is simulated using the HSPICE platform[51] which is interfaced with MATLAB in order to input the circuit simulation parameters required as well as to evaluate the output obtained from circuit simulation. The CMOS latches are designed using the 10nm PTM model[52]; a 5-fin FinFET design is used in order to achieve the desired drive strength for each inverter. The FeFETs are implemented using a circuit compatible SPICE model which tracks the history-dependent switching behavior of the ferroelectric and the details of the model have been presented in our prior work[50]. Additionally, we also consider the role of variation as well as the wire parasitics. A nominal variation of 10 mV in threshold voltage (per fin) is considered for the CMOS latches (effect of increasing $V_T$ variation is detailed in supplementary note 3). Additionally, for the interconnect routing in the FeFET array, a line-to-ground capacitance of 0.122 fF/μm & 0.109 fF/μm, and a line-to-line capacitance of 0.0229 fF/μm & 0.0217 fF/μm is considered for the two metal layers- M1 and M2, required, respectively[44]. A random noise source, available in the spice stimulus[53], of maximum amplitude 50mV is added to the supply voltage.

Computing the MaxCut using the above array is a two-step process, and is illustrated here with the aid of a small 4-node graph as shown in Fig. 5c: *(a) Programming the FeFET array to represent the input graph:* During this phase, the latches are decoupled from the FeFET array by turning OFF the interfacing switches. The FeFET array can now be programmed as a standard memory array, and we employ the half-select ($V_W/2$) bias scheme (Fig. 5b) to write the desired state to the FeFETs. First, a negative write voltage $V_W$ (-3 V considered here) pulse is applied to all the WLs ensuring that all the FeFETs are initialized to the high-$V_T$ (low conductance) state. Subsequently, a column-wise write scheme is used wherein all the FeFET cells corresponding to $A_{ij}$=1 (in the respective column) are programmed into the low-$V_T$ (high conductance) state; no programming is required for the FeFETs that represent to $A_{ij}$=0 since the whole FeFET array was initialized into the high $V_T$ (low conductance) state at the start, as discussed earlier. Programming the FeFETs into the low-$V_T$ (high conductance) state is achieved by asserting the word line to +3V and the corresponding BL and SL to 0V. The column-wise programming scheme as well as the applied voltages are selected such that the half-selected cells in the corresponding column and row experience minimal program disturbs (see supplementary note 4).

*(b) Execution phase:* Once the FeFET array has been programmed to represent the input graph, the latches are powered on, and connected to the rows and columns of the FeFET array. The FeFETs now act as coupling elements among the latches. Subsequently, the coupled system evolves towards the ground-state energy- manifested as certain latches changing their state, and in the process, computing the solution to the MaxCut problem. Considering the statistical nature of the computation, wherein the system can get trapped in local minima (resulting in a sub-optimal solutions), the power supply to the latches and the enable signal to the switches interfacing the latches and the FeFET array are cycled five times.

Fig. 5d shows the programming and execution of a 4-node representative problem using the array proposed above. Once the FeFET array has been programmed according to the adjacency matrix of the graph, the interface between the latch and the array is turned ON and OFF periodically five times. It can be observed that, each time the latches are connected to the array, the latch outputs evolve towards the ground-state energy, and the final state ($V_{out1}= V_{out3}=V_{DD}$; $V_{out2}= V_{out4}=0$) represents the two sets of the created by the MaxCut of the input graph (Fig. 5d). We note that optimal solutions are observed in all five evaluations here. However, this may not be the case in all cases, particularly, as the graph size increases.

Subsequently, we evaluate the dynamics of the proposed array using numerous (30) graph instances of varying sizes up to 50 nodes; 10 randomly instantiated graphs are considered for each graph size. The entire two-stage process including the programming of the FeFETs is simulated in each case, and hence the simulations are computationally intensive. We also observe that in all the evaluated instances, the latches settle to a steady-state within 100 ns after the transmission gates are turned ON. Fig. 5e shows the MaxCut computed by the proposed array (normalized to the optimal solution). We observe that for smaller graphs (10 nodes) optimal solutions are obtained in 9 out of the 10 cases. However, the solution quality degrades (mean accuracy for the 50 node graphs is 94%) as the graph size is increased owing to the increasing complexity of the solution space- a feature observed in all Ising machine implementations.

Next, we also evaluate the energy efficiency of the proposed design (calculated as solutions per Watt per second); the energy and time required for both the stages of operation (i.e., programming the FeFETs and the execution of the MaxCut problem) are considered. Fig. 6 shows the energy efficiency with increasing problem size. For 50 node graphs (largest size simulated here), we compute an energy efficiency of $5 \times 10^8$ solutions per Watt per second, and a projected efficiency of $1.04 \times 10^8$ solutions per Watt per second for 100 node problems (assuming connectivity of 50% and a settling time of 100 ns); the projection for 100 node graphs is performed in order to facilitate comparison with other works. Fig. 6b compares the observed efficiency with other approaches and confirms that the proposed system not only provides the advantage of CMOS compatibility but is also competitive in terms of energy efficiency.

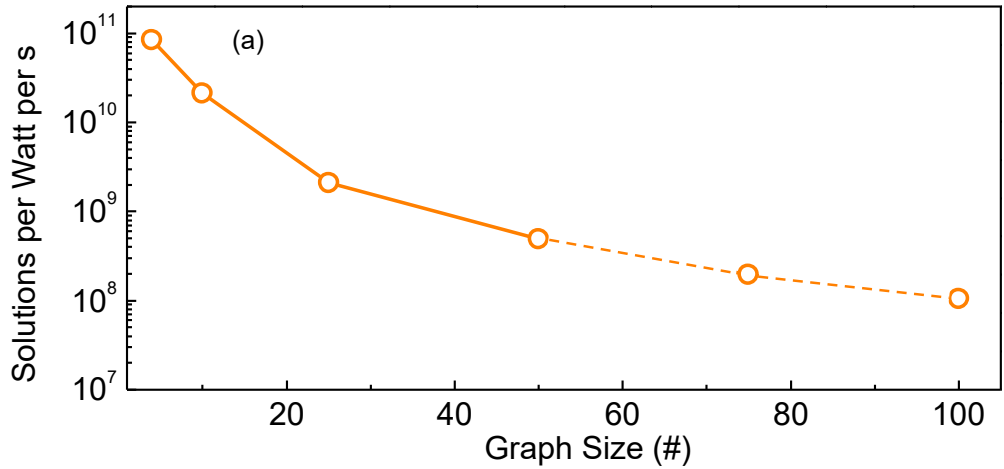

| (b) | This work | Quantum Annealing [25] | Coherent Ising Machine (CIM) [26] | Coupled oscillators [41][42][44] | | BRIM [47] |
|---|---|---|---|---|---|---|
| Ising Spin | CMOS Latch | Qubit | DOPO (degenerate parametric oscillators) | Ring oscillator (RO) / Schmitt Trigger oscillator | Phase transition (IMT)-based oscillators (e.g., $VO_2$) | ZIV diode |
| Implementation | Cross-coupled inverters (**4T design**) | Josephson Junction | Laser Pulses | -RO: 2N+1 inverter stages (min. 6T design) -**Schmitt trigger:** 6T CMOS design + RC feedback | 1T + 1R design R: phase transition material | OPAMP + Resistors (multi-transistor design required for OPAMP) |
| Coupling | FeFET | Flux Storage | Digital feedback (FPGA) | Resistive/ capacitive | Resistive / capacitive | Resistive |
| CMOS Compatibility | Yes | Needs special processing | Need special processing | Yes | May require special processing | Yes |
| Potential Challenge | Variability | Cryogenic operation | Power, Scalability | Variability | Variability, temperature stability | Scalability |
| Operating Temperature | Room temperature | 0.015 K (Cryogenic Temperature) | Room temperature | Room temperature | Room temperature | Room temperature |
| Energy Efficiency (solutions/W/s) | $1.04 \times 10^8$ | $< 4 \times 10^{-9}$ | - | - | $1.3 \times 10^7$ | - |

**Fig. 6. Projected Energy efficiency of Latch-based Ising machine.** (a) Evolution of energy efficiency (solutions per Watt per second) as a function of graph size; 50% connectivity is assumed. (b) Comparison of the proposed Latch-based Ising machine with other design approaches.

**Discussion**

In summary, the proposed implementation provides a pathway to realizing a compact and scalable Ising machine to solve computationally challenging problems such as MaxCut using CMOS-compatible components. Additionally, the energy efficiency of the CMOS latch and ferroelectric FETs facilitate high energy efficiency with a projected capability of computing >$10^8$ solutions per joule of energy. Finally, the design is also well-positioned to take advantage of the maturity of the CMOS process technology to realize scaled implementations, making it a promising design approach for realizing high-performance application-specific accelerators for solving combinatorial optimization problems.

**Methods:**

**Device fabrication**

FeFETs employed in this work have a poly-crystalline Si/TiN/doped $HfO_2$/$SiO_2$/p-Si gate stack, which are integrated on the 28 nm node high-κ metal gate CMOS technology platform on 300 mm silicon wafers. Detailed information is described elsewhere[50]. For the ferroelectric gate stack, a thin $SiO_2$ interfacial layer is grown first, followed by the deposition of the doped $HfO_2$ film. Then a TiN metal gate electrode was deposited using physical vapor deposition, following that the poly-Si gate electrode is deposited. The source and drain n+ regions were formed by phosphorous ion implantation then a rapid thermal annealing at ~1000 °C. This step also results in the formation of the ferroelectric orthorhombic phase within the doped $HfO_2$.

**Data Availability**

The datasets generated during and/or analyzed during the current study are available from the corresponding author on reasonable request.

**Code Availability**

All codes used in this work are either publicly available or available from the authors upon reasonable request.

**Acknowledgment:**

This work was supported in part by NSF ASCENT grant (No. 2132918). It is also partially supported in part by the Army Research Office under Grant Number W911NF-21-1-0341. We also acknowledge GlobalFoundries, Dresden Germany for providing the testing devices.

**Author contributions**

A.M and N.S developed the main idea. K. N, A.A and V.N participated in discussions and further development of the idea. A.M and Z.Z performed the experiments. Y.X and Y.X developed the FeFET model. A.M, M.K.B, S.A, M.M.I performed the simulations. A.M, K.N and N.S wrote the manuscript. All authors discussed the results and commented on the manuscript.

**Competing interests**

The authors declare no competing interests.

**Supplementary Information**

**CMOS-Compatible Ising Machines built using Bistable Latches Coupled through Ferroelectric Transistor Arrays**

Antik Mallick[1], Zijian Zhao[2], Mohammad Khairul Bashar[1], Shamiul Alam[3], Md Mazharul Islam[3], Yi Xiao[4], Yixin Xu[4], Ahmedullah Aziz[3], Vijaykrishnan Narayanan[4], Kai Ni[2], Nikhil Shukla[1]*

[1]Department of Electrical and Computer Engineering, University of Virginia, Charlottesville, VA 22904, USA

[2]Department of Electrical and Microelectronic Engineering, Rochester Institute of Technology, Rochester, NY 14623, USA.

[3] Department of Electrical Engineering and Computer Science, University of Tennessee, Knoxville, TN 37996, USA

[4]Department of Computer Science & Engineering, Pennsylvania State University, State College, PA 16801, USA

*email: ns6pf@virginia.edu

**Supplementary Figures**

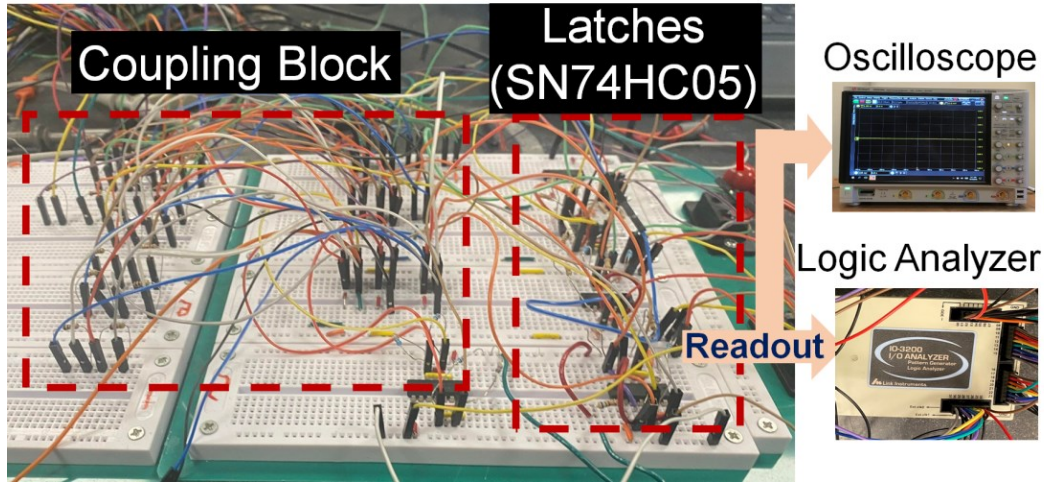

**Supplementary Figure 1| Picture of the experimental setup.** The latches are implemented using IC74HC05. The outputs are measured using an oscilloscope and logic analyzer.

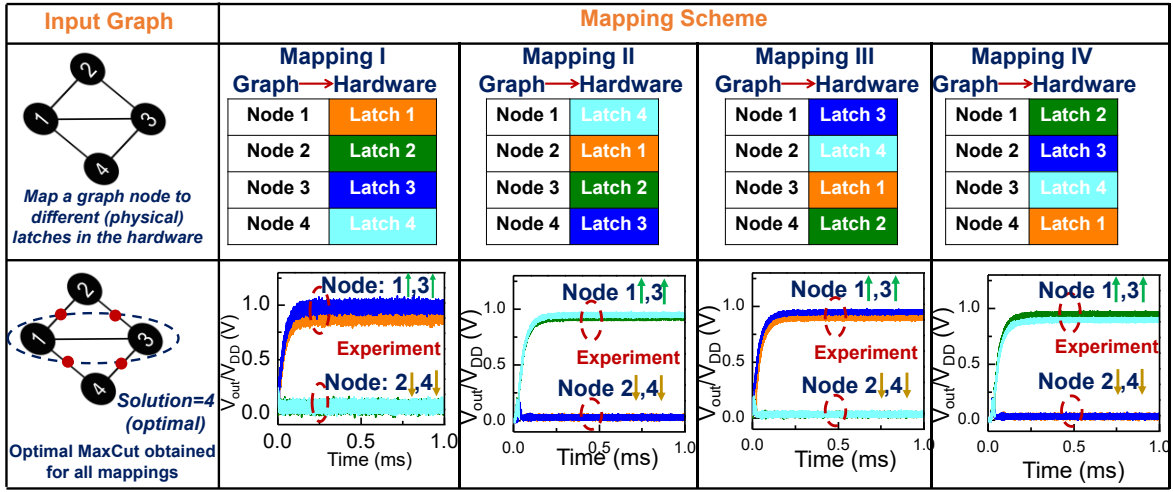

**Supplementary Figure 2| Evaluation of the impact of asymmetry among the CMOS Latches.** Observed MaxCut solution when the nodes in the input graph are mapped to different physical latches in the hardware. Optimal solution is observed for all the mappings indicating that the observed behavior is not the result of inherent asymmetry among the latches.

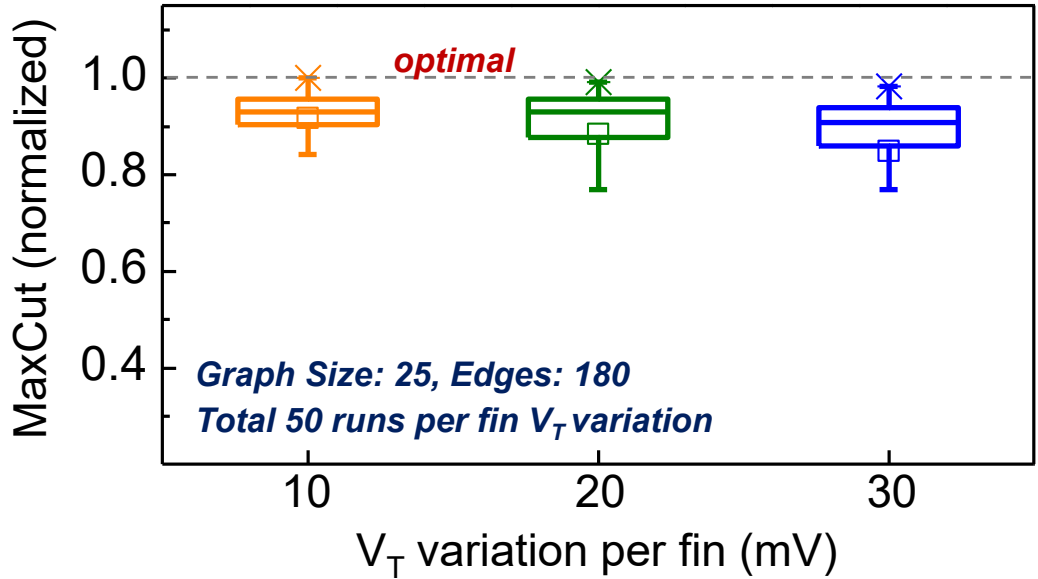

**Supplementary Figure 3| Effect of Threshold Voltage Variation.** Simulation showing the effect of threshold voltage variation (per fin) in the latches on the MaxCut solution computed by the latch based Ising machine array; Monte Carlo simulations (3σ, 50 runs) are performed for each case. Box plot results; center: median; box: interquartile range (IQR); whiskers: 1.5× IQR)

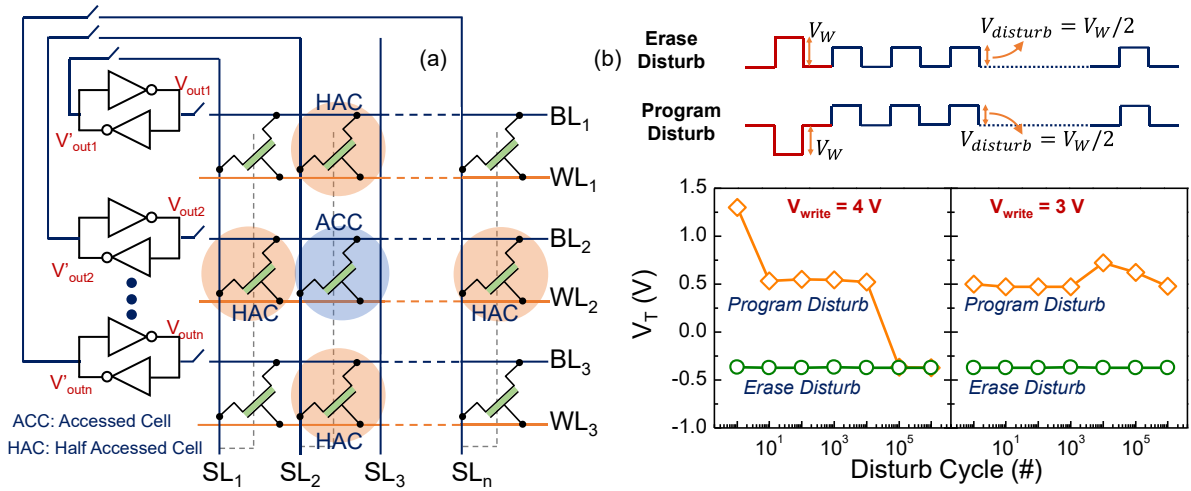

**Supplementary Figure 4| Write Disturb in FeFET Array.** (a) Schematic of the 1T-FeFET used in the Ising machine implementation. The (blue) circled cells indicate an accessed cell while the (orange) circled cells indicate the half-accessed cells. (b) Evolution of the threshold voltage in the HAC cells corresponding to repetitive program and erase of the ACC cell for two different write voltages, $V_W$= 4V

# Supplementary Notes

## Supplementary Note 1. Experimental Setup

Supplementary figure 1 shows the experimental setup for measuring the dynamics of resistively coupled latches as an Ising machine. The latch-based Ising spin is implemented using discrete CMOS inverters (IC SN74HC05N), which are subsequently, coupled through external resistors that behave as coupling elements. The outputs are measured using a digital oscilloscope (Keithley DSOS104A) and a logic analyzer (Link Instruments IO-3200). The inverters are biased using a supply voltage of 5V sourced through a Keithley 2400 source meter.

## Supplementary Note 2. Evaluation of Asymmetry among Latches

To rule out the possibility that the observed output states of the latches are caused by inherent asymmetry (arising from variation) amongst them, we map the nodes in the illustrative graph to different physical latches, as shown in supplementary figure 2. For instance, in mapping I, nodes 1, 2, 3, 4 are mapped to latches 1, 2, 3, 4, respectively, and in mapping II, nodes 1, 2, 3, 4 are mapped to latches 4, 1, 2, 3, respectively. It can be observed that the latches settle to optimal MaxCut solution irrespective of the mapping, indicating that the latch asymmetry does not dominate the outputs of the latches in the experiment.

**Supplementary Note 3. Effect of Threshold Voltage Variation in FeFET coupled Latches**

We explore the effect of threshold voltage ($V_T$) variation (expressed threshold variation per fin) in the transistors (used for the latches) on the behavior of the proposed Ising machine array (Fig. 5 of main text). The latches are constructed using 10nm PTM models (5-fin FinFET design is used), and the analysis is performed using a Monte Carlo approach (50 runs). It is observed from supplementary figure 3 that the system is able to compute high-quality MaxCut solutions in the presence of variation although the mean accuracy of the solution reduces with increasing $V_T$ variation. Further, the simulations indicate that $V_T$ variation >20 mV variation per fin can have a significant impact on the solution quality.

**Supplementary Note 4. Write Disturb in FeFET Array**

We program the FeFET array as a standard memory array by adopting the half-select ($V_W/2$) write scheme wherein a column-wise write operation is performed with a write voltage ($V_W$) of 3V (Fig. 5 of main text). The reason behind this choice of write voltage was to ensure that the half-accessed cells experience minimum write disturb i.e., unintentional change in threshold voltage is response to programming of other cells. It can be observed from supplementary figure 4b that the half-accessed cells show minimal disturb when $V_W$= 3V even after $10^4$ cycles; in contrast, a write voltage of 4V induces significant disturb which would have an adverse impact on the programming operation.